\def\la{\mathrel{\hbox{\rlap{\hbox{\lower4pt\hbox{$\sim$}}}\hbox{$<$}}}}
\def\ga{\mathrel{\hbox{\rlap{\hbox{\lower4pt\hbox{$\sim$}}}\hbox{$>$}}}}

\def\Teff{\ifmmode{T_{\rm eff}}\else{\hbox{$T_{\rm eff}$} }\fi}
\def\Rzero{\ifmmode{R_0}\else{\hbox{$R_0$} }\fi}

\def\56ni{$^{56}$Ni}
\def\56co{S^{56}$Co}

\def\v10{$v_{10}$(S\lowercase{i}~II)}
\def\ref{\reference{}}

\def\T{SN~1991T--like}

\def\bg{SN~1991bg--like}

\def\86G{SN~1986G--like}

\def\aa{SN~1999aa--like}

\documentclass{aastex}

\received{}
\accepted{}
\journalid{}{}
\articleid{}{}
\shortauthors{Hatano et al.}
\shorttitle{Type Ia SN 1994D}

\begin{document}
\title {Spectroscopically Peculiar Type Ia Supernovae and
Implications for Progenitors}

\author {David Branch}

\affil{Department of Physics and Astronomy, University of
Oklahoma, Norman, Oklahoma 73019}

\begin{abstract}

In a recent paper Li et~al. (2000) reported that 36\% of 45 Type~Ia
supernovae discovered since 1997 in two volume--limited supernova
searches were spectroscopically peculiar, and they suggested that
because this peculiarity rate is higher than that reported for an
earlier observational sample by Branch et~al. (1993), it is now more
likely that SNe~Ia are produced by more than one kind of progenitor.
In this paper I discuss and clarify the differences between the
results of Li et~al. and Branch et al. and I suggest that multiple
progenitor systems are now less likely than they were before.

\end{abstract}


\section{Introduction}

In a recent paper Li et~al. (2000; hereafter Li00) reported that 36\%
of 45 Type~Ia supernovae (SNe~Ia) discovered since 1997 in the Lick
Observatory Supernova Search and the Beijing Astronomical Observatory
Supernova Search were spectroscopically peculiar.  They concluded that
because this peculiarity rate is higher than the $13$ to $17$\%
peculiarity rate reported for an earlier observational sample by
Branch, Fisher, \& Nugent (1993; hereafter BFN), it is now more likely
than it was before that SNe~Ia are produced by more than one kind of
progenitor (e.g., both single--degenerate and double--degenerate
binary systems).  My purpose in this paper is to discuss and clarify
the differences between the results of Li00 and BFN and to suggest
that the results of Li00, together with some other recent results,
make multiple progenitor systems less likely than before.

\section{The BFN and Li00 Results}

\subsection{Old Results: BFN}

BFN defined a normal SN~Ia to be one whose optical spectra resemble
those of SNe~1981B, 1989B, 1992A, and 1972E.  [For plots of spectra
and references to the original observational papers see BFN, and for a
good general review of supernova spectra see Filippenko (1997).] Near
the time of maximum light normal SNe~Ia have deep absorption features
near 6150~\AA\ due to Si~II $\lambda6355$ and near 3750~\AA\ due to
Ca~II $\lambda3945$, and other features that are produced by lines of
O~I, Mg~II, Si~II, S~II, Ca~II, and Co~II.  For weeks after maximum
the Si~II 6150~\AA\ and Ca~II 3750~\AA\ absorptions remain strong, as
Fe~II lines make their appearance.

A spectroscopically peculiar SN~Ia was defined to be one that has
different spectral features (not just different expansion velocities).
At that time there were three well observed examples: SNe~1991T,
1991bg, and 1986G.  Near the time of maximum light SN~1991T was
strikingly peculiar in having prominent features of Fe~III and hardly
any signs of Si~II, S~II, and Ca~II. (The spectrum obtained closest to
maximum light actually was obtained three days before maximum.)  By a
week after maximum the usual Si~II, S~II, and Ca~II features had
developed but they were weaker than in a normal SN~Ia. By three weeks
after maximum the spectrum of SN~1991T looked almost normal.

The most conspicuous peculiarity of SN~1991bg was the presence of a
broad absorption trough extending from about 4150 to 4400~\AA, due to
a blend of Ti~II lines.  The spectroscopic peculiarities of SN~1986G
were like those of SN~1991bg but less extreme.

BFN inspected all available spectra of SNe~Ia and attempted to
subclassify them as either normal, or like SN~1991T, or SN~1991bg, or
SN~1986G. (In this paper, for consistency with Li00, events like
SN~1991bg and 1986G will be counted together and referred to as
SN~1991bg--like.)  For 83 events BFN offered a least a partial
subclassification, i.e., for some events it was possible to exclude
some but not all of the peculiar categories.  For example, for 12
events the SN~1991bg category could be excluded, but because the first
spectrum was obtained too long after maximum light and/or the quality
of the spectrum was insufficient, the SN~1991T category was not ruled
out.  Such events were not counted as normal.  Of the 45 events for
which BFN did offer a full subclassification, 39 (89\%) were normal
and five (11\%) were peculiar.  (The peculiar events were the three
prototypes plus SNe~1957A and 1960H, both of which were found to be
like \bg.)  If three additional events that were suspected of
peculiarity were counted as peculiar, then 83\% were normal and 17\%
were peculiar.  (These three were SN~1988G, suspected of being
\T, and SNe 1980I and 1971I, both suspected of being \bg.)

BFN assumed that the observational sample that they were working with
(whatever was available) was more like a magnitude--limited sample
than like a volume--limited one, so that it was likely to be strongly
biased against the subluminous \bg\ events.  They suggested that \bg\
events are perhaps not all that rare compared to the bright,
observationally conspicuous events that are called normal.

\subsection{New Results: Li00}

The results of Li00 are based on intensive supernova searches
characterized by such small time baselines and deep limiting
magnitudes that, as shown by Li, Filippenko, \& Riess (2000), they are
essentially volume--limited searches as far as SNe~Ia are concerned.
This means that Li00 were able to probe the true fraction of the
SN~1991bg--like events.  Of the 45 events in their sample, Li00 found
7 to be SN~1991bg--like, compared to 4 to 6 such events in the BFN
sample.  Evidently the \bg\ events are not as common as BFN expected.
In hindsight, the reason may be that the BFN sample was not so much
like a magnitude--limited sample after all.  In a genuinely
magnitude--limited sample the volume within which a supernova of
absolute magnitude $M$ can be detected depends strongly on $M$, being
proportional to $10^{-0.6 M}$.  However, many of the SNe~Ia in the BFN
sample were found in targeted searches of nearby galaxies, where the
discovery probability probably depended much less strongly on $M$.

Li00 found just one definite \T\ event, SN~1997br, the extensive
observations of which were presented and discussed by Li
et~al. (1999).  However, Li00 also found something quite new and
interesting --- SN~1999aa.  Li00 showed that one day before the time
of maximum light the spectrum of SN~1999aa was that of a normal SN~Ia,
but seven days before maximum it resembled SN~1991T in having
conspicuous Fe~III features and a weak Si~II 6150\AA\ absorption while
at the same time resembling normal SNe~Ia in having a fairly strong
Ca~II 3750\AA\ absorption.  Li00 found 6 \aa\ events (Li00 designated
them as SN~1991T$_{aa}$ and counted them as \T\ events but for what
follows in this paper they must be considered separately) and two
events that could have been either \T\ or \aa.  Clearly there is no
conflict between the 1 to 3 genuinely \T\ events in the Li00 sample
and the 1 to 2 such events in the BFN sample.  The one
significant difference between the results of Li00 and BFN is that
Li00 found from 6 to 8 \aa\ events while BFN did not identify any.

Li00 suggested that they found a higher SN~1991T--like fraction than
BFN because of an age bias, i.e., if the first spectrum was
obtained too long after maximum light a SN~1991T--like peculiarity
would have been difficult to see, so the SN~Ia was erroneously
subclassified as normal.  In principle, this age bias should not apply
to \T\ events in the BFN sample because if BFN thought that the first
spectrum was too late to show a SN~1991T peculiarity they did not
subclassify it as normal. And, as we have just seen, there is no
conflict between Li00 and BFN about \T\ events that needs to be
explained.

However, the age bias certainly applied to \aa\ events in the BFN
sample (see below).  In fact, as Li00 discuss, even their results are
affected by an age bias against \aa\ events.  According to Table~1 of
Li00, 19 of the 29 SNe~Ia that they subclassified as normal were first
observed spectroscopically near or after the time of maximum
brightness.  If \aa\ events are to be counted as peculiar, these 19
events should not have been subclassified as normal because some of
them may have been \aa.  Among those events for which a \aa\
peculiarity could have been detected --- e.g., those observed at least
six days before maximum, 11 were normal, 5 were \aa, 3 were
\bg\, and one was \T, i.e, 45\% were peculiar!  (And ``peculiar''
begins to sound like an inappropriate term.)

\subsection{The BFN Sample Revisited}

In view of the Li00 results I looked again at the BFN sample.  First,
although there is no conflict between the results of Li00 and BFN for
\T\ events, I checked on whether the age bias could have been a
significant factor for
\T\ events.  In the BFN sample 39 events were subclassified as normal.
For about two thirds of these there is clear evidence that they were
not \T.  For some of the others, especially those that were not
observed spectroscopically before a week after maximum brightness, the
case that they were not \T, although reasonably convincing to BFN,
might not be convincing to others.  (In a few cases, e.g., SNe~1981F
and 1988B, the evidence is not now convincing to me.)  Thus the age
bias may have had a significant effect on the BFN results in the sense
that if fewer events had been subclassified as normal, the peculiarity
fraction would have come out to be higher.

For the \aa\ events the age bias was much more important.  Only 5 of
the 39 SNe~Ia that BFN subclassified as normal were observed as early
as six days before the time of maximum light --- SNe~1974G, 1984A,
1989B, 1990N, and 1992A.  All 5 of these were normal, none were
\aa.  Between the times of the BFN
sample, which extended only to SN~1992A, and the Li00 sample, which
began with SN~1997Y, SNe~1994D (Patat et~al. 1996; Meikle et~al. 1996;
Filippenko 1997) and 1996X (Salvo et~al. 2000) also were observed well
before maximum light, and they too were normal, not \aa.  The
difference between this and the fact that among the SNe~Ia of Li00
that were observed six or more days before maximum, there were 5 \aa\
events versus 11 normal events, can only be attributed to
small--number statistics.  Many more good pre--maximum spectra of
SNe~Ia are needed, to find out just how common the \aa\ events really
are, and to determine whether or not there is a continuous range of
spectral characteristics among SNe~Ia at such early times.

\section{Implications for Progenitors}

The issue of whether SNe~Ia are produced by single--degenerate or
double--degenerate binary systems, or both, is still open, and the
issue of whether SNe~Ia come from carbon ignitors or helium ignitors,
or both, is perhaps still not closed. [For explanations of these terms
and a review of SN~Ia progenitor candidates see, e.g, Branch et
~al. (1995)].  Li00 argued that the higher the peculiarity rate of
SNe~Ia, the more likely it is that multiple progenitors produce
SNe~Ia.  This is a significant conclusion because it may tend to erode
astronomers' confidence in using SNe~Ia as distance indicators at high
redshift (the progenitor mix would evolve) and it could even have a
bearing on the success or failure of proposals for more ambitious
searches for high--redshift SNe~Ia.

I suggest that the results of Li00 do not make multiple progenitors
more likely now than they were before.  First, in a general sense, the
sheer relative numbers of peculiar events don't tell us whether
multiple progenitors are involved or not --- only inferences from the
observations about the physical links, or lack thereof, between the
various SN~Ia subclasses can do that.  More specifically, the fact
that a genuinely volume--limited sample contains a higher fraction of
\bg\ events than the BFN sample wasexpected, so it has no new
implications for multiple progenitors.  And as we have seen, the
fraction of \T\ events in the Li00 and BFN samples are similar.  The
one important difference between the Li00 and BFN results is the
discovery by Li00 of the \aa\ events --- and if, as mentioned by Li00,
they appear be a missing link between normal SNe~Ia and \T\ events,
then the \aa\ events make multiple progenitors seem less likely than
before.

Physical considerations do make the \aa\ events appear to be unifying.
The spectroscopic differences among SNe~Ia at early times are caused
primarily by temperature differences (Mazzali et~al. 1993; Nugent
et~al. 1995; Hatano et~al. 1999), with abundance differences playing a
secondary role (and thus being difficult to establish). A sufficiently
low temperature leads to Ti~II lines.  A sufficiently high temperature
leads to Fe~III lines and the weakness or absence of Si II and Ca II
lines.  The \aa\ events, which have both Fe~III and Ca II lines a week
before maximum light and normal spectra by the time of maximum light,
appear to connect the \T\ events with normal SNe~Ia.

Another recent development makes it less likely that SN~1991T was
physically distinct from normal SNe~Ia.  Fisher et~al. (1999) found
that SN~1991T was too luminous to have been produced by a
Chandrasekhar--mass explosion, and therefore suggested that it must
have been a super--Chandrasekhar product of a double--degenerate
progenitor.  Fisher et~al. assumed that the distance to NGC~4527, the
parent galaxy of SN~1991T, is $16.4\pm1.0$~Mpc, on the basis of
Cepheid--derived distances to NGC~4536 and NGC~4496A, which appear to
be members of the same small group of galaxies as NGC~4527.  However,
from a simple model of the polarized dust echo of SN~1991T Sparks et
al. (1999) have estimated an upper limit to NGC~4527 of 15~Mpc, and
from Cepheids in NGC~4527 Saha et~al. (2000) find $14.0\pm0.9$~Mpc.
The reduction in the distance to NGC~4527, if correct, leaves SN~1991T
somewhat overluminous for a SN~Ia but it weakens the argument for
super--Chandrasekhar mass ejection and a double--degenerate
progenitor.  Whether a \T\ event in the Hubble flow will prove to be
too luminous for Chandrasekhar mass ejection remains to be seen.

Other recent developments have favored the view that most and perhaps
all SNe~Ia are carbon ignitors in single--degenerate systems, with
Chandrasekhar mass ejection [see Livio (2000) and Nomoto (2000) for
reviews; Langer et~al. (2000), Khokhlov (2000), and Ruiz--Lapuente
\& Canal (2000) for further
important developments; and Branch (2001) for a recent review].  In
this case the observational diversity of SNe~Ia would be caused
primarily by differences in the ejected mass of $^{56}$Ni (which
controls the temperature and the peak luminosity), secondarily to
differences in the amount of mass that is ejected at high velocity
[differences that may indicate that two modes of burning propagation
--- deflagrations and delayed detonations --- are involved (Hatano et
al. 2000; Lentz et~al. 2001)], and to a lesser extent differences in
progenitor metallicity (Lentz et~al. 2000) and perhaps some other
things.

I thank Weidong Li and all members of the University of Oklahoma
supernova group for discussions and Abi Saha for communicating a
result in advance of publication.  This work was supported by NSF
grant AST--9986965 and NASA grant NAG5-3505.

\clearpage

\begin {references}

\ref Branch, D. 2001, in Young Supernova Remnants, ed. S.~S.~Holtz and
U.~Hwang (AIP), in press

\ref Branch, D., Fisher, A., \& Nugent, P. 1993, AJ, 106, 2383 (BFN)

\ref Branch, D., Livio, M., Yungelson, L. R., Boffi,~F., \&
Baron,~E. 1995, PASP, 107, 1019

\ref Filippenko, A. V. 1997, ARAA, 35, 309

\ref Fisher, A., Branch, D., Hatano, K., \& Baron, E., 1999, MNRAS, 304,
67

\ref Hatano, K., Branch,~D., Fisher,~A., Millard,~J., \& Baron,~E.
1999, ApJS, 121, 233

\ref Hatano, K., Branch, D., Lentz, E., Baron, E., Filippenko,~A.~V.,
\& Garnavich,~P. 2000, ApJ, 543, L49

\ref Khokhlov, A. 2000, ApJ, submitted (astro-ph/0008463)

\ref Langer, N., Deutschmann, A., Wellstein, S., \&
H\"oflich,~P. 2000, ApJ, in press (astro-ph/0008444)

\ref Lentz, E., Baron, E., Branch, D., Hauschildt,~P., \&
Nugent,~P. 2000, ApJ, 530, 966

\ref Lentz, E., Baron, E., Branch, D., \& Hauschildt, P. 2001
ApJ, in press (astro-ph/0007302)

\ref Li, W. D. et al., 1999, AJ, 117, 2709

\ref Li, W. D., Filippenko, A. V., Treffers, R. R., Riess,~A.~G.,
Hu,~J., \& Qiu,~Y. 2000, ApJ, in press (Li00) (astro-ph/0006292)

\ref Li, W. D., Filippenko, A. V., \& Riess, A. G. 2000, ApJ, in press (astro-ph/0006291)

\ref Livio, M. 2000, in The Greatest Explosions Since the Big Bang:
Supernovae and Gamma Ray Bursts, ed. M.~Livio, N.~Panagia, \& K.~Sahu
(Cambridge University Press), in press (astro-ph/0005344)

\ref Mazzali, P. A., Lucy, L. B., Danziger, I. J., Guiffes,~C.,
Cappellaro,~E., \& Turatto,~M. 1993, A\&A, 279, 447

\ref Meikle, W. P. S. et~al. 1996, MNRAS, 281, 263

\ref Nomoto, K., Umeda, H., Kobayashi, C., Hachisu,~I, \&
Tsujimoto,~T. 2000, in Cosmic Explosions, ed. S.~S.~Holt \&
W.~W.~Zhang (AIP), in press (astro-ph/0003134)

\ref Nugent, P. E., Phillips, M. M., Baron, E., Branch,~D., \&
Hauschildt,~P. 1995, ApJ, 455, L147

\ref Patat, F., Benetti, S., Cappellaro, E., Danziger,~I.~J., Della
Valle,~M., Mazzali,~P., \& Turatto,~M. 1996, MNRAS, 278, 111

\ref Ruiz--Lapuente, P. \& Canal, R. 2000, ApJL, submitted (astro-ph/0009312)

\ref Saha, A., Sandage, A., Thim, F., Tammann, G. A., Labhardt,~L.,
Christensen,~J., Macchetto,~F.~D., \& Panagia,~N. 2000, preprint

\ref Salvo, M. E., Cappellaro,~E., Mazzali,~P.~A., Benetti,~S.,
Danziger,~I.~J., Patat,~F., \& Turatto,~M. 2000, MNRAS, in press
(astro-ph/0009065)

\ref Sparks, W. B., Macchetto, F. D., Panagia, N., Boffi,~F.~R.,
Branch,~D., Hazen,~M., \& Della~Valle,~M. 1999, ApJ, 523, 585

\end{references}

\end{document}